\documentclass[a4paper,11pt]{article}
\usepackage{amssymb}
\usepackage{amsfonts}
\usepackage{graphicx}
\usepackage{dcolumn}
\usepackage{amsmath}
\usepackage{latexsym,bm}
\usepackage{geometry}

\def \be {\begin{equation}}
\def \ee {\end{equation}}
\def \bea {\begin{eqnarray}}
\def \eea {\end{eqnarray}}
\def \nn {\nonumber}

\def \a {\alpha}
\def \b {\beta}

\def \d {\delta}

\def \m {\mu}
\def \n {\nu}
\def \k {\kappa}

\def \s {\sigma}
\def \r {\rho}
\def \o {\omega}

\def \th {\theta}
\def \Th {\Theta}

\def \t {\tau}
\def \dag {\dagger}
\def \p {\partial}

\def\bd{\begin{document}}
\def\ed{\end{document}}
\def\nn{\nonumber}
\def\bea{\begin{eqnarray}}
\def\eea{\end{eqnarray}}
\let\bm=\bibitem
\let\la=\label

\def\N{{\cal N}}
\def\sst{\scriptscriptstyle}
\def\thetabar{\bar\theta}
\def\Tr{{\rm Tr}}
\def\one{\mbox{1 \kern-.59em {\rm l}}}

\def\a{\alpha}      \def\da{{\dot\alpha}}
\def\b{\beta}       \def\db{{\dot\beta}}
\def\c{\gamma}  \def\C{\Gamma}  \def\cdt{\dot\gamma}
\def\d{\delta}  \def\D{\Delta}  \def\ddt{\dot\delta}
\def\e{\epsilon}        \def\vare{\varepsilon}
\def\f{\phi}    \def\F{\Phi}    \def\vvf{\f}
\def\h{\eta}
\def\k{\kappa}
\def\l{\lambda} \def\L{\Lambda}
\def\m{\mu} \def\n{\nu}
\def\o{\omega}
\def\P{\Pi}
\def\r{\rho}
\def\s{\sigma}  \def\S{\Sigma}
\def\t{\tau}
\def\th{\theta} \def\Th{\Theta} \def\vth{\vartheta}
\def\X{\Xeta}
\def\z{\zeta}
\def\w{\wedge}
\def\u{\underline}
\def\hs{\hspace}

\def\cA{{\cal A}} \def\cB{{\cal B}} \def\cC{{\cal C}}
\def\cD{{\cal D}} \def\cE{{\cal E}} \def\cF{{\cal F}}
\def\cG{{\cal G}} \def\cH{{\cal H}} \def\cI{{\cal I}}
\def\cJ{{\cal J}} \def\cK{{\cal K}} \def\cL{{\cal L}}
\def\cM{{\cal M}} \def\cN{{\cal N}} \def\cO{{\cal O}}
\def\cP{{\cal P}} \def\cQ{{\cal Q}} \def\cR{{\cal R}}
\def\cS{{\cal S}} \def\cT{{\cal T}} \def\cU{{\cal U}}
\def\cV{{\cal V}} \def\cW{{\cal W}} \def\cX{{\cal X}}
\def\cY{{\cal Y}} \def\cZ{{\cal Z}}

\def\ua{\underline{\alpha}} \def\ubb{\underline{\beta}}
\def\ug{\underline{\gamma}}
\def\ub{\underline{\phantom{\alpha}}\!\!\!\beta}
\def\uc{\underline{\phantom{\alpha}}\!\!\!\gamma}
\def\um{\underline{\mu}} \def\un{\underline{\nu}}
\def\ud{\underline\delta}
\def\ue{\underline\epsilon}
\def\una{\underline a}\def\unA{\underline A}
\def\unb{\underline b}\def\unB{\underline B}
\def\unc{\underline c}\def\unC{\underline C}
\def\und{\underline d}\def\unD{\underline D}
\def\une{\underline e}\def\unE{\underline E}
\def\unf{\underline{\phantom{e}}\!\!\!\! f}\def\unF{\underline F}
\def\unm{\underline m}\def\unM{\underline M}
\def\unn{\underline n}\def\unN{\underline N}
\def\unp{\underline{\phantom{a}}\!\!\! p}\def\unP{\underline P}
\def\unq{\underline{\phantom{a}}\!\!\! q}
\def\unQ{\underline{\phantom{A}}\!\!\!\! Q}
\def\unH{\underline{H}}
\def\ul{\underline}

\def\As {{A \hspace{-6.4pt} \slash}\;}
\def\bs {{b \hspace{-6.4pt} \slash}\;}
\def\Ds {{D \hspace{-6.4pt} \slash}\;}
\def\ds {{\del \hspace{-6.4pt} \slash}\;}
\def\ss {{\s \hspace{-6.4pt} \slash}\;}
\def\ks {{ k \hspace{-6.4pt} \slash}\;}
\def\ps {{p \hspace{-6.4pt} \slash}\;}
\def\pas {{{p_1} \hspace{-6.4pt} \slash}\;}
\def\pbs {{{p_2} \hspace{-6.4pt} \slash}\;}

\def\Fh{\hat{F}}
\def\Vh{\hat{V}}
\def\Xh{\hat{X}}
\def\ah{\hat{a}}
\def\xh{\hat{x}}
\def\yh{\hat{y}}
\def\ph{\hat{p}}
\def\xih{\hat{\xi}}

\def\psit{\tilde{\psi}}
\def\Psit{\tilde{\Psi}}
\def\tht{\tilde{\th}}

\def\At{\tilde{A}}
\def\Qt{\tilde{Q}}
\def\Rt{\tilde{R}}
\def\Nt{\tilde{N}}

\def\at{\tilde{a}}
\def\st{\tilde{s}}
\def\ft{\tilde{f}}
\def\pt{\tilde{p}}
\def\qt{\tilde{q}}
\def\vt{\tilde{v}}
\def\nt{\tilde{n}}

\def\delb{\bar{\partial}}
\def\bz{\bar{z}}
\def\bD{\bar{D}}
\def\bB{\bar{B}}

\def\bk{{\bf k}}
\def\bl{{\bf l}}
\def\bp{{\bf p}}
\def\bq{{\bf q}}
\def\br{{\bf r}}
\def\bx{{\bf x}}
\def\by{{\bf y}}
\def\bR{{\bf R}}
\def\bV{{\bf V}}

\def\d{\delta}\def\D{\Delta}\def\ddt{\dot\delta}

\def\p{\partial} \def\del{\partial}
\def\xx{\times}
\def\uno{\mbox{1 \kern-.59em {\rm l}}}

\def\trp{^{\top}}
\def\inv{^{-1}}
\def\dag{{^{\dagger}}}

\def\pr{\prime}
\def\rar{\rightarrow}
\def\lar{\leftarrow}
\def\lrar{\leftrightarrow}

\title{Black holes in Truncated Higher Spin AdS$_3$ Gravity}
\author{
Bin Chen$^{1,2,3}$\footnote{bchen01@pku.edu.cn},\,
Jiang Long$^{1,2}$\footnote{lj301@pku.edu.cn}\,
and
Yinan Wang$^{1}$\footnote{ynwang@pku.edu.cn}
}
\date{}

\begin{document}

\maketitle

\begin{center}
{\it
$^{1}$Department of Physics, Peking University, Beijing 100871, P.R. China\\
\vspace{2mm}
$^{2}$State Key Laboratory of Nuclear Physics and Technology, Peking University, Beijing 100871, P.R. China\\
\vspace{2mm}
$^{3}$Center for High Energy Physics, Peking University, Beijing 100871, P.R. China\\
}
\vspace{10mm}
\end{center}

\begin{abstract}
We study the higher spin black holes in a truncated version of higher spin gravity in $AdS_3$. This theory contains only finite number of even spins $s=2,4,\cdots,2N$. 
We mainly focus on the simplest case, so-called (Type I and II) spin ${\tilde 4}$ gravity, which contains only spin 2 and spin 4 fields. This spin ${\tilde 4}$ gravity is as simple as spin 3 gravity, thus  provides another example to test various ideas on higher spin gravity. We find that the asymptotical symmetry of this spin ${\tilde 4}$ gravity is a classical ${W}(2,4)$-symmetry. Moreover, we study the black hole solution with pure spin 4 hair and discuss its thermodynamics. One important feature of this black hole is that its entropy could be written in compact forms. Furthermore, we investigate a $G_2$ generated higher spin gravity. This higher spin gravity only contains spin 2 and spin 6 fields which makes it different from other kinds of higher spin gravity. We find the corresponding black hole with spin 6 hair, and discuss its thermodynamics analytically. 
 \end{abstract}

\newpage
\section{Introduction}

Higher spin gauge theory in $AdS_4$, which was constructed decades ago\cite{Vasiliev I, Vasiliev II, Vasiliev III}, has attracted much attention in the context of AdS/CFT correspondence in recent years. While this theory originally describes interacting higher spins $s=0,1,2,3,\cdots$, it allows for a minimal truncation to a theory with only even spins. This minimal bosonic Vasiliev theory(A-type) was conjectured to be dual to 3D free or critical $O(N)$ vector model\cite{Klevanov}, depending on the bulk scalar boundary conditions. This duality has been analyzed and extended in various aspects in the past few years. For the complete references, see a nice review \cite{Giombi:2012he}.

On the other hand, the higher spin theories become even simpler in $AdS_3$.  The pure higher spin theory without scalar in $AdS_3$ can be written as the summation of two Chern-Simons terms with opposite level\cite{Blencowe:1989, Blencowe:1990}, much like the pure gravity case\cite{Townsend:1986,Witten:1988}. While an one-parameter theory with higher spin algebra $hs[\lambda]$ describes an infinite tower of higher spins in general, a theory with finite number of higher spins is also well-defined for the choice of $\lambda=3,4,5,\cdots$. The physical spectrum depends on the choice of the gauge group and the embedding of $sl(2,R)$. In an interacting theory with spins from $2$ up to $N$, the gauge group of Chern-Simons action is $sl(N,R)$. More interestingly, in \cite{ASG I:2010, ASG II:2010} the asymptotically symmetry algebras were analyzed for $AdS_3$ higher spin theories,  which turns out be $\mathcal{W}$ algebras.\footnote{See \cite{ASG 3, ASG 4, ASG 5,ASG 6} for other discussions and various generalizations on asymptotic symmetries.} Just like the original Brown-Henneaux analysis\cite{Brown:1986} on $AdS_3$ gravity, this suggests that there exists a two dimensional conformal field theory with $\mathcal{W}$ symmetry dual to the higher spin gravity. In the infinite spin limit, it was conjectured in \cite{Gaberdial:2011} that the dual CFT is a coset $WZW$ minimal model. The evidence for this conjecture is based on symmetry, spectrum matching and RG flow considerations, much like the Higher Spin/O(N) conjecture in \cite{Klevanov}. For further developments see a recent review\cite{Gaberdial:2012}. A truncated version with even spins of this conjecture has been considered in \cite{Ah, Gaberdial2:2011}.

In another route, the $AdS_3$ higher spin black holes were firstly considered in \cite{Per kraus:2011,Ammon:2011nk}. This treatment of higher spin black holes was guided by the dual conformal field theory. The higher spin charge and its corresponding chemical potential were recognized by OPE and Ward identity analysis. The most important lesson from the study is that a gauge invariant quantity (holonomy) encodes the information of thermodynamics of higher spin black holes, while conventional notions such as the diffeomorphism and the horizons should be revised. More explicitly, it was required that the holonomy around the time circle should give the same eigenvalues as they are in the BTZ black holes\cite{BTZ1, BTZ2}. Such requirement was shown to give consistent thermodynamics\cite{Per kraus:2011, Theisen:2012}\footnote{For a canonical approach to this problem, see \cite{tempo:2012}.}. For the simplest example, the spin 3 black hole, this leads to a compact expression of entropy, even though a generalized Cardy formula has not been fully understood. However, for other examples including $\mathcal{W}_{\infty}[\lambda]$ black holes \cite{kraus:2011,Matthias:2012} and spin 4 black hole\cite{Hai:2011}, the discussion of black hole thermodynamics could only be restricted to the perturbation level. See \cite{HSBH} for a recent review. Due to the mixing of spin 2 and higher spin gauge transformation, an independent but very valuable understanding of higher spin black holes from physical metric and higher spin fields points of view is still lack so far. It seems that a second order formalism of higher spin gravity may shed light on this problem. See \cite{Campoleoni:2012hp, Ippei Fujisawa} for recent investigations on this issue.

Note that there are two kinds of truncations in $AdS_3$ higher spin gravity. The first kind is to truncate the infinite higher spins with $s=1,2,3,\cdots$ to even spins $s=2,4,\cdots$. The second kind is to truncate the infinite higher spins to finite number of higher spins with $s=1,2,\cdots,N'$. It is interesting to investigate if a combination of these two kinds of truncation can give us a well-defined theory with only finite number of even spins $s=2,4,\cdots,2N$. This turns out to be the case.  It can be easily seen from the $hs[\lambda]$ algebra. Firstly,  choosing even spin generators $V^s_m$, the reduced algebra which we denote by $\widetilde{hs}[\lambda]$ is obviously closed. Second, setting the parameter $\l$ to be integer  and moding out the ideal the resulting theory would only contains finite number of even spins. Note that in principle we may obtain two classes of truncated higher spin gravity. When $\lambda$ is even, we denote it as Type I Spin $\widetilde{2N}$ gravity. When $\lambda$ is odd, we denote it as Type II Spin $\widetilde{2N}$ gravity.   With $2N$, it remind us that the theory contains only even spins from spin 2 up to spin $2N$, and with $``\sim''$ it distinguish from usual $sl(N,R)$ higher spin gravity. Actually they are the even spin truncations of $sl(N,R)$ higher spin gravity. And it could be considered to be simpler than conventional $sl(N,R)$ higher spin gravity as it involves less fields up to a finite spin.

In these Spin $\widetilde{2N}$ theory, the spin 4 would be the lowest higher spin 
and hence it is in principle as simple as the spin 3 field in conventional $sl(3,R)$ gravity. This may provide another tractable model to test all kinds of ideas on higher spin gravity. Actually as locally $sp(4,R)\simeq so(5,R)$, two types of Spin $\tilde{4}$ gravity are quite similar and share the same asymptotical $W(2,4)$ symmetry. 
In both types of Spin $\tilde{4}$ gravity, the black holes with spin 4 hair can be constructed analytically, with their entropy functions having pleasant simple forms. They are much like the spin 3 black holes considered previously, and provide the second type of examples that one can explore the full thermodynamics of higher spin black holes. 



The success of finding the black holes with pure spin 4 hair, whose thermodynamics could be studied analytically, inspires us to search for other black holes with a single higher spin hair. It turns out that the only other case is the black hole with spin 6 hair in a higher spin gravity based on  the lowest exceptional group\footnote{Actually, it is the noncompact version of $G_2$. However, this distinction is not quite important in practice.} $G_2$. In this $G_2$ gravity, the spectrum only contains spin 2 and spin 6 higher spin fields. Such a higher spin gravity can be obtained by truncate the $hs[\lambda]$ theory to Type II Spin $\tilde{6}$ gravity, and then truncate out the spin 4 generators from Type II Spin $\tilde{6}$ gravity.  Besides this exotic feature, the other properties of this higher spin gravity are quite similar to others. Since it contains only one higher spin hair, the thermodynamics of $G_2$ higher spin black holes can be studied analytically. Hence $G_2$ gravity provides the third prototype of higher spin gravity.

 In the next section, we give a brief introduction to AdS$_3$ high spin gravity and its finite truncations. In Section 3, we investigate the Spin $\tilde{4}$ gravity. We discuss and give the corresponding asymptotically symmetry algebra ${W}(2,4)$  explicitly. We construct spin $\tilde{4}$ black hole in Type I Spin $\tilde{4}$ gravity  and explore its  thermodynamics  analytically. In section 4, we discuss the $G_2$ gravity and find that the corresponding asymptotically symmetry algebra is ${W}(2,6)$. The thermodynamics of the $G_2$ black hole is also included. In section 5, we discuss the black holes with spin 4 and spin 6 hairs in type I and type II spin $\tilde{6}$ gravity. In section 6, we end with conclusion and discussions. In appendices, we collect the conventions in our study.  



\section{Higher Spin Gravity in $AdS_3$ and Truncations}

The action of higher spin gravity in $AdS_3$ can be written as a sum of two Chern-Simons terms with opposite level,
\be
S_{HSG}=S_{CS}[A]-S_{CS}[\bar{A}],\label{CS}
\ee
where
\be
S_{CS}[A]=\frac{k}{4\pi}\int tr(A dA+\frac{2}{3}A^3)
\ee
and we omit the wedge for brevity. $A$ and $\bar{A}$ are one forms and take value in suitable Lie algebra. For example, taking the gauge group to be $sl(N,R)\times sl(N,R)$ will describe a tower of higher spin $s=2,3,\cdots,N$ with each higher spin appear once\footnote{In this paper, we extensively use the Principal Embedding.}. The parameter $k$ in this action is called the level and it is related to the Newton constant $G$ and $AdS_3$ radius $l$ by the relation
\be
k=\frac{l}{4G}.
\ee
For simplicity, we set the radius $l$ to be 1 from now on. To relate the theory to gravity, we need to introduce vielbein $e$ and spin connection $\omega$ and relate them to the gauge potential by
\be
A=\omega+e,\ \bar{A}=\omega-e.
\ee
The field equation from the above action turns out to be
\be
F\equiv dA+A\wedge A=0,\ \bar{F}\equiv d\bar{A}+\bar{A}\wedge\bar{A}=0.\label{EOM}
\ee
The gauge transformations are
\be
\delta A=d\Lambda+[A,\Lambda],\ \delta\bar{A}=d\bar\Lambda+[\bar{A},\bar{\Lambda}].
\ee
The infinite extension of $sl(N,R)$ will be the so-called one parameter higher spin algebra $hs[\lambda]$ and the corresponding higher spin gravity will contain a infinite tower of higher spin with $s=2,3,\cdots,\infty$.

 Let us start from the infinite dimensional $hs[\lambda]$ algebra and do truncation. The notation we use are the same as the one in \cite{HSBH} and \cite{ASG 6}. The algebra consists of generators $V^s_m$ for every $s$ and $-(s-1)\leq m\leq (s-1)$. They are generated by a $sl(2,\mathbb{R})$ subalgebra with generators $L_{-1},L_0,L_1$, which obey usual commutation relations:
\begin{equation}
[L_1,L_0]=L_1,\hs{3ex}[L_{-1},L_0]=-L_{-1},\hs{3ex}[L_1,L_{-1}]=2L_0.
\end{equation}
Then we let $V^s_{s-1}=L_1^{s-1}$ for every $s$, and $V^s_{m-1}=-[L_{-1},V^s_m]/(s+m-1)$ for every $m$. Note in this notation we have $V^2_1=L_1,V^2_0=L_0,V^2_{-1}=L_{-1}$. The commutation rule between $V^s_m$ and $V^t_n$ is:
\begin{equation}
[V^s_m,V^t_n]=\sum_{u=2,4,6\dots}^{s+t-|s-t|-1} g^{st}_u(m,n;\lambda)V^{s+t-u}_{m+n}\label{Commutation}
\end{equation}
where
\begin{equation}
g^{st}_u(m,n;\lambda)=\frac{q^{u-2}\phi^{st}_u(\lambda)N^{st}_u(m,n)}{2(u-1)!}.
\end{equation}
The expressions of $\phi^{st}_u(\lambda)$ and $N^{st}_u(m,n)$ are:
\begin{align}
&\phi^{st}_u(\lambda)= _4 F_3\left[\begin{array}{c|c}\frac{1}{2}+\lambda,\frac{1}{2}-\lambda,\frac{2-u}{2},\frac{1-u}{2}&1\\ \frac{3}{2}-s,\frac{3}{2}-t,\frac{1}{2}+s+t-u\end{array}\right]\\
&N^{st}_u(m,n)=\sum_{k=1}^{u-1}(-1)^k \binom{u-1}{k} [s-1+m]_{u-1-k} [s-1-m]_k [t-1+n]_k [t-1-n]_{u-1-k}\nn
\end{align}
and $q$ is a parameter that can be tuned for different conventions.
Here $_4 F_3$ is hypergeometric function, its general definition is:
\begin{equation}
_p F_q\left[\begin{array}{c|c}a_1,a_2,\dots,a_p&z\\ b_1,b_2,\dots,b_q\end{array}\right]=\sum_{n=0}^\infty\frac{[a_1]_n\dots[a_p]_n}{[b_1]_n\dots[b_q]_n}\frac{z^n}{n!},
\end{equation}
in which  the Pochhammer symbol has been used
\begin{equation}
[a]_n=a(a-1)\dots(a-n+1).
\end{equation}
Generally the coefficients are complicated, but for $s=2$ case, we have a simple equation:
\begin{equation}
[V^2_m,V^t_n]=(m(t-1)-n)V^t_{m+n}.
\end{equation}

For different $\lambda$ we have inequivalent algebras. If $\lambda=N$ is an integer, an ideal  consisting of all the $s>N$ generators forms. If one mod out this ideal, the algebra reduces to $sl(N,\mathbb{R})$.

Note in the commutation relation (\ref{Commutation}) if we assume $s$ and $t$ to be even numbers, the right side has only even components. So a subalgebra with only even generators exists, we call it\footnote{In \cite{Gaberdial2:2011}, it was called $hs[\l]^{(e)}$.} $\widetilde{hs}(\lambda)$. If $\lambda=2N$ is an even number,
this algebra can be truncated to an finite $sp(2N,R)$ algebra and the resulting theory is called Type I Spin $\widetilde{2N}$ gravity . If $\l=2N+1$ is an odd number, the algebra can be truncated to a noncompact real form of $so(2N+1,R)$ algebra and the corresponding theory is the
Type II Spin $\widetilde{2N}$ gravity.   

For example, if $N=2$, we have Type I Spin $\tilde{4}$ gravity and Type II Spin $\tilde{4}$ gravity. The former can be obtained from the usual spin 4 gravity. For the usual spin 4 gravity, it is constructed by taking gauge group to be $sl(4,R)$ and it has 15 generators which we denote by $\{L_{-1},L_0,L_1,W^3_{-2},\cdots,W^3_2,W^4_{-3},\cdots,W^4_3\}$. It is obvious that the subset $\{L_{-1},L_0,L_1,W^4_{-3},\cdots,W^4_3\}$ is a subalgebra $sp(4,R)$ of $sl(4,R)$ and the commutation relations are given in the Appendix A. The latter can be obtained from the usual spin 5 gravity. For the usual spin 5 gravity, it is constructed by taking gauge group to be $sl(5,R)$ and it has 24 generators which we denote by $\{L_{-1},L_0,L_1,W^3_{-2}, \cdots, W^3_2, W^4_{-3}\cdots,W^4_3,\\W^5_{-4},\cdots,W^5_4\}$. Also, the subset $\{L_{-1},L_0,L_1,W^4_{-3},\cdots,W^4_3\}$ forms $so(5,R)$ which is a subalgebra of $sl(5,R)$.  Such a spin $\tilde{4}$ gravity is as simple as the spin 3 gravity. It will be the main focus of the following discussion.


\section{Spin $\tilde{4}$ gravity}

In this section, we study the properties of  two types of Spin $\tilde{4}$ gravities, which are based on $sp(4,R)$ and $so(5,R)$ algebras respectively. Due to the fact that $sp(4,R)$ is locally isomorphic to $so(5,R)$, two types of theories are quite similar, and have the same asymptotical symmetry algebra. Here we mainly work on the Type I Spin $\tilde{4}$ gravity.



\subsection{Asymptotical Symmetry}

To be definite, we focus on Type I Spin $\tilde{4}$ gravity. Since the analysis is standard and  is explained explicitly in \cite{ASG II:2010}, we just quickly sketch the main points.
Upon making gauge transformation, one can show that the solution of (\ref{EOM}) can be cast into the form\cite{ASG II:2010}\footnote{The ``bar'' term can be treated similarly.}
\be
A=b^{-1}(a(x^+)+d)b.
\ee
with $b=\exp{\rho (L_0)}$. And asymptotically $AdS_3$ can be represented as
\be
a=(L_1+\frac{2\pi}{k} \mathcal{L}(x^+)L_{-1}+\frac{5\pi}{9k} \mathcal{W}(x^+)W^4_{-3})dx^+\label{*}.
\ee
By identifying the transformation that preserve asymptotically $AdS_3$ boundary condition and use the definition of Poisson bracket from Chern-Simons action, one can find that the asymptotically symmetry algebra is
\bea
\{\mathcal{L}(\theta),\mathcal{L}(\theta')\}&=&-(\delta(\theta-\theta')\mathcal{L}(\theta)'+2\delta'(\theta-\theta')\mathcal{L}(\theta))-\frac{k}{4\pi}\delta(\theta-\theta')'''\nonumber\\
\{\mathcal{L}(\theta),\mathcal{W}(\theta')\}&=&-(3\delta(\theta-\theta')\mathcal{W}(\theta)'+4\delta'(\theta-\theta')\mathcal{W}(\theta))\nonumber\\
\{\mathcal{W}(\theta),\mathcal{W}(\theta')\}&=&-\frac{k}{400\pi}(\delta^{(7)}(\theta-\theta')+\frac{112\pi}{k}\mathcal{L}(\theta)\delta^{(5)}(\th-\th')+\frac{280\pi}{k}\mathcal{L}(\th)'\delta^{(4)}(\th-\th')+\nonumber\\
 & &+U_3(\th)\delta^{(3)}(\th-\th')+U_2(\th)\delta''(\th-\th')+U_1(\th)\delta'(\th-\th')+U_0(\th))
\eea
where the $U_i$ are defined as
\bea
U_0&=&\frac{12\pi}{k}\mathcal{L}^{(5)}+\frac{2832\pi^2}{k^2}\mathcal{L}'\mathcal{L}''+\frac{1248\pi^2}{k^2}\mathcal{L}\mathcal{L}'''+\frac{27648\pi^3}{k^3}\mathcal{L}^2\mathcal{L}'+\frac{40\pi}{k}\mathcal{W}'''+\frac{2240\pi^2}{k^2}\mathcal{L}\mathcal{W}'\nonumber\\
& &+\frac{2240\pi^2}{k^2}\mathcal{L}'\mathcal{W}\nonumber\\
U_1&=&\frac{80\pi}{k}\mathcal{L}^{(4)}+\frac{5632\pi^2}{k^2}\mathcal{L}\mathcal{L}''+\frac{4720\pi^2}{k^2}\mathcal{L}'^2+\frac{18432\pi^3}{k^3}\mathcal{L}^3+\frac{200\pi}{k}\mathcal{W}''+\frac{4480\pi^2}{k^2}\mathcal{L}\mathcal{W}\nonumber\\
U_2&=&\frac{9408\pi^2}{k^2}\mathcal{L}\mathcal{L}'+\frac{224\pi}{k}\mathcal{L}'''+\frac{360\pi}{k}\mathcal{W}'\nonumber\\
U_3&=&\frac{3136\pi^2}{k^2}\mathcal{L}^2+\frac{336\pi}{k}\mathcal{L}''+\frac{240\pi}{k}\mathcal{W}
\eea
This algebra is a classical $W(2,4)$ algebra\cite{Lukyanov:1989gg,Lukyanov:1990tf,W symmetry} with central charge $c=6k=\frac{3l}{2G}$. Correspondingly, we may expect that the dual CFT should have the same W-symmetry. 
One candidate with such symmetry is the coset CFT $so(4)_k\times so(4)_1/so(4)_{k+1}$\cite{Gaberdial2:2011}. However, this CFT is not unitary  in the limit of large $c$\footnote{We would like to thank anonymous referee for pointing out this possibility.}.

\subsection{Type I Spin $\tilde{4}$ Black Holes}

This subsection is to explore the thermodynamics of black holes with spin 4 hair in Type I Spin $\tilde{4}$ gravity in AdS$_3$. In the previous work\cite{Per kraus:2011,Ammon:2011nk}, it has been shown that the holonomy condition can give consistent  thermodynamics of higher spin black holes. 
Consider, in the boundary CFT of spin $\tilde{4}$ gravity, a deformation is reached by adding  an irrelevant operator $\int \mu \mathcal{W}+\bar{\mu} \bar{\mathcal{W}}$ to the action\footnote{A subscript 4 is omitted since there is only one kind of higher spin charge and chemical potential. }. This will deform the UV structure of boundary CFT. In the bulk, this corresponds to add a source term in the
ansatz of gauge field component $A_-$ and $\bar{A}_+$. Let us use the standard gauge fixed form
\be
A=b^{-1}a b+b^{-1}db, \hs{3ex}\bar{A}=b\bar{a}b^{-1}+bdb^{-1}.\label{Gauge}
\ee
where
\be
a=(L_1-\mathcal{L}L_{-1}+\mathcal{W} W^4_{-3})dx^++(\nu L_{-1}+\sum_{m=-3}^{m=3} q_{m} W^4_m)dx^-\label{ansatz}
 \ee
 and similar form for $\bar{a}$. In this ansatz we have twist a sign of $\mathcal{L}$ related to (\ref{*}) and made some other redefinitions which are evident. The spin 4 black hole with nonzero spin 3 charge has been extensively studied in \cite{Hai:2011} so we will skip the details and go to the new result. The flat condition can be found to be equal to
\be
da+a\w a=0
\ee
Note that to require the coefficients in $a$ are constants reduces the previous equation to
\be
[a_+,a_-]=0
\ee
Solve the equations and one can see that
\bea
q_{-3} = \frac{1}{5}(-5 \mathcal{L}^3 q_{3} - 11 \mathcal{L} \mathcal{W} q_{3}),\\
 q_{-2} = 0,\hs{3ex} \nu = \frac{54 \mathcal{W} q_{3}}{5}
  ,\\ q_{-1} =
 3 (\mathcal{L}^2 q_{3} + \mathcal{W} q_{3}),\\
  q_{1} = -3 \mathcal{L} q_{3}, \hs{3ex}q_{0} = 0,\hs{3ex} q_{2} = 0.
  \eea
  From the Ward identity one can interpret $q_{3}=\mu$, which is the chemical potential for spin-4 charge. Note that $\mu$ is related to the potential $\alpha$, which appears in the partition function
\be
  q_{3}=\mu=\frac{\alpha}{\bar{\tau}}.\label{q3}
  \ee 
  The holonomy can be written as
  \be
 \omega=2\pi (\tau a_+-\bar{\tau}a_-).\label{omega}
 \ee
 As has been shown in \cite{Per kraus:2011}, we should impose the condition that the eigenvalues of the honolomy are the same as the BTZ case. In the Type I Spin $\tilde{4}$ case, one finds that the eigenvalues of the holonomy are
 \be
 \pm i\pi, \hs{3ex}\pm i3\pi
 \ee
 Hence we should require
 \be
 P_2=tr{\omega^2}=-20\pi^2; \hs{3ex}P_3=tr{\omega^3}=0; \hs{3ex}P_4=tr{\omega^4}=164\pi^4.
 \ee
Note that in our case, the second condition is satisfied automatically. This just shows that a consistency truncation from spin 4 HS gravity can be possible. To see whether this gives a consistent thermodynamics, one should show that there exist a entropy function $S(\mathcal{L},\mathcal{W})$ such that
\bea
\tau=i\frac{c_1}{4\pi^2}\frac{\partial S}{\partial \mathcal{L}},\\
\alpha=i\frac{c_2}{4\pi^2}\frac{\partial S}{\partial \mathcal{W}}
\eea
where the constants $c_1, c_2$ are fixed by the consideration of $\mathcal{T}\mathcal{W}$ OPE and Ward identity. From the discussion in \cite{Hai:2011}, one can see that a suitable choice of $c_1,c_2$ can be
\be
c_1=\frac{2\pi}{k}, \hs{3ex}c_2=\frac{5}{18}c_1.
\ee
The entropy should be equal to the BTZ entropy when the spin 4 charge $\mathcal{W}=0$. In addition, from dimensional consideration one finds that a suitable choice of entropy can be
\be
S(\mathcal{L},\mathcal{W})=2\pi k\sqrt{\mathcal{L}}f(y)\label{entropy}
\ee
where $y=\frac{\mathcal{W}}{\mathcal{L}^2}$. Taking all conditions into account, one finds that the holonomy equations reduce to the following two equations
\bea
 1&=&f(y)^2 + \frac{4}{9} (4 + 7 y - 36 y^2)f'(y)^2,\label{a}\\
 164\pi^4&=&-\frac{4}{81} \pi^4 (81 (-41 + 36 y) f(y)^4 -
   1296 (-4 - 7 y + 36 y^2) f(y)^3
f'(y) \nn\\
&&+ 1944 (1 + 4 y)^2 (-4 + 9 y) f(y)^2
f'(y)^2 - 576 (4 + 7 y - 36 y^2)^2 f(y)
f'(y)^3 \nn\\
& & + 16 (-41 + 36 y) (4 + 7 y - 36 y^2)^2
f'(y)^4).
\eea
The initial condition should be $f(0)=1$. Then from the first equation one finds that
\be
f(y)=\cos(\theta(y))\label{**}
\ee
with
\be
\theta(y)=\frac{1}{4} (\arcsin\frac{7}{25} - \arcsin\frac{7 - 72 y}{25}).\label{theta}
\ee
One can check that this solution satisfy the second equation as well.  Substituting the ansatz (\ref{**}) into the first equation (\ref{a}), one finds that $-\frac{1}{4}\leq y\leq \frac{4}{9}$ so the extremal black hole is at $y=\frac{4}{9}$ or $y=-\frac{1}{4}$.
Note that our truncation of spin 4 higher spin gravity to spin $\tilde{4}$ makes life simpler as spin 3 higher spin gravity. The truncation to arbitrary spin $\widetilde{2N}$ gravity is straightforward but it is hard to find a simple black hole solutions on which the holonomy condition could be solved analytically.

Let us have a closer look at the solution (\ref{**}). It is much like the original spin 3 black hole solution. However, there is a remarkable difference. For spin 3 solution in \cite{Per kraus:2011}, $y\sim \frac{\mathcal{W}_3^2}{\mathcal{L}^3}$, so the entropy is invariant under $\mathcal{W}_3\to-\mathcal{W}_3$. However, our solution of spin $\tilde{4}$ goes like $y\sim\frac{\mathcal{W}_4}{\mathcal{L}^2}$, hence it is not invariant under $\mathcal{W}_4\to-\mathcal{W}_4$. We believe that this charge-conjugation violation will be a general feature in spin $\widetilde{2N}$ gravity and it may be traced to the distinction between even and odd spins. More interestingly, this violation leads to the fact that there are four kinds of extreme black holes for fixed $\mathcal{L},\bar{\mathcal{L}}$ and their entropies are not equal. It remains to have a deeper understanding of such kind of phenomenon. There could be other kinds of solutions which dominate the thermodynamics as discussed in \cite{David:2012iu}. Hence the extremal solutions we found above are not actually thermodynamically favored.

Before we end our discussion on Spin $\tilde{4}$ gravity, we would like to comment briefly on the
spin $\tilde{4}$ black hole in Type II Spin $\tilde{4}$ Gravity. Since Type II Spin $\tilde{4}$ gravity is truncated from the $sl(5,R)$ gravity, while Type I Spin $\tilde{4}$ gravity is truncated from $sl(4,R)$ gravity, one can easily see that the holonomy of the black holes in these two spin $\tilde{4}$ gravity theories are obviously different. However, the entropy function turns out to be of the similar form. We will not go to the details.

\section{$G_2$ Gravity}

From the previous discussion and the result of spin 3 gravity, we find that the simplest higher spin black holes are the ones that contain only one kind of higher spin hair. So far the existing examples include the spin 3 black hole and the spin 4 black holes in Type I,  II Spin $\tilde{4}$ gravity above. Hence it is interesting to search for other examples that contain a single higher spin hair. The guiding principle is that if such kind of theory exist, the Lie algebra should have rank two. This restricts the possible algebra to be $A_2, B_2, C_2, D_2$ and $G_2$. While $A_2$ corresponds to spin 3 gravity, and $B_2(\simeq C_2)$ corresponds to Type I(II) Spin $\tilde{4}$ gravity, $D_2$ has 6 generators which cannot have higher spin $s>2$ hair, the only candidate algebra is the exceptional $G_2$.
 A matrix representation of $G_2$ algebra and their commutation relations are given in Appendix B. It has 14 generators and a suitable embedding of $sl(2,R)$ gives us that the spectrum contains spin 2 and spin 6 fields. The $G_2$ gravity is based on the exceptional group $G_2$. Since the study is almost the same as before, we just list the result below.
The asymptotically symmetry algebra is\footnote{The $G_2$ generated W algebra can be found in \cite{Fesher}} ${W}(2,6)$ with central charge $c=6k$:
\begin{align}
&\{\mathcal{L}(\theta),\mathcal{L}(\theta')\}=-(\delta(\theta-\theta')\mathcal{L}(\theta)'+
2\delta'(\theta-\theta')\mathcal{L}(\theta))-\frac{k}{4\pi}\delta(\theta-\theta')'''\nonumber\\
&\{\mathcal{L}(\theta),\mathcal{W}(\theta')\}=-(5\delta(\theta-\theta')\mathcal{W}(\theta)'+6\delta'(\theta-\theta')\mathcal{W}(\theta))\nonumber\\
&\{\mathcal{W}(\theta),\mathcal{W}(\theta')\}=-(\frac{k\delta^{(11)}(\theta-\theta')}{2352\pi}+\frac{55}{294}\mathcal{L}\delta^{(9)}(\theta-\theta')+
\frac{165}{196}\mathcal{L}'\delta^{(8)}(\theta-\theta')\nonumber\\&+\left(\frac{1364\pi\mathcal{L}^2}{49k}+\frac{99\mathcal{L}''}{49}\right)\delta^{(7)}(\theta-\theta')
+\left(\frac{1364\pi}{7}\mathcal{L}\mathcal{L}'+\frac{22}{7}\mathcal{L}'''\right)\delta^{(6)}(\theta-\theta')+V_5(\theta)\delta^{(5)}(\theta-\theta')+\nonumber\\
&V_4(\theta)\delta^{(4)}(\theta-\theta')+V_3(\theta)\delta^{(3)}(\theta-\theta')V_2(\theta)\delta^{(2)}(\theta-\theta')+V_1(\theta)\delta^{(1)}(\theta-\theta')+V_0(\theta))\nonumber\\
\end{align}
where we have omitted a index `6' to simplify notation and hope this does not cause confusion. The functions in the third Poisson bracket are
\begin{align}
&V_5=\frac{244640\pi^2\mathcal{L}^3}{147k^2}-\frac{26}{21}\mathcal{W}+\frac{14355\pi\mathcal{L}'^2}{49k}+\frac{17160\pi\mathcal{L}\mathcal{L}''}{49k}+
\frac{165}{49}\mathcal{L}^{(4)}\nonumber\\
&V_4=\frac{611600\pi^2\mathcal{L}^2\mathcal{L}'}{49k^2}-\frac{65}{21}\mathcal{W}'+\frac{43065\pi\mathcal{L}\mathcal{L}''}{49k}
+\frac{19030\pi\mathcal{L}\mathcal{L}'''}{49k}+\frac{495}{196}\mathcal{L}^{(5)}\nonumber\\
&V_3=\frac{5395456\pi^3\mathcal{L}^4}{147k^3}-\frac{2480}{21k}\pi\mathcal{L}\mathcal{W}+\frac{3677960\pi^2\mathcal{L}\mathcal{L}'^2}{147k^2}+
\frac{2195072\pi^2\mathcal{L}^2\mathcal{L}''}{147k^2}\nonumber\\&+\frac{528\pi\mathcal{L}''^2}{k}-\frac{10}{3}\mathcal{W}''+
\frac{16390\pi\mathcal{L}'\mathcal{L}'''}{21k}+\frac{40700\pi\mathcal{L}\mathcal{L}^{(4)}}{147k}+\frac{55}{42}\mathcal{L}^{(6)}\nonumber\\
&V_2=\frac{10790912\pi^3\mathcal{L}^3\mathcal{L}'}{49k^3}-\frac{1240\pi\mathcal{W}\mathcal{L}'}{7k}+\frac{615780\pi^2\mathcal{L}'^3}{49k^2}-
\frac{1240\pi\mathcal{L}\mathcal{W}'}{7k}+\frac{44968\pi^2\mathcal{L}\mathcal{L}'\mathcal{L}''}{k^2}\nonumber\\
&+\frac{485936\pi^2\mathcal{L}^2\mathcal{L}^{(3)}}{49k^2}+\frac{704\pi\mathcal{L}''\mathcal{L}^{(3)}}{k}-\frac{40\mathcal{W}^{(3)}}{21}+
\frac{20460\pi\mathcal{L}'\mathcal{L}^{(4)}}{49k}+\frac{6094\pi\mathcal{L}\mathcal{L}^{(5)}}{49k}+\frac{22\mathcal{L}^{(7)}}{49}\nonumber\\
&V_1=\frac{9830400\pi^4\mathcal{L}^5}{49k^4}-\frac{36608\pi^2\mathcal{L}^2\mathcal{W}}{21k^2}+\frac{48668416\pi^3\mathcal{L}^2\mathcal{L}'^2}{147k^3}
-\frac{3470\pi\mathcal{L}'\mathcal{W}'}{21k}+\frac{19336192\pi^3\mathcal{L}^3\mathcal{L}''}{147k^3}\nonumber\\
&-\frac{736\pi\mathcal{W}\mathcal{L}'^2}{7k}+\frac{1108404\pi^2\mathcal{L}'^2\mathcal{L}''}{49k^2}+\frac{660864\pi^2\mathcal{L}\mathcal{L}''^2}{49k^2}-
\frac{2000\pi\mathcal{L}\mathcal{W}''}{21k}+\frac{2929544\pi^2\mathcal{L}\mathcal{L}'\mathcal{L}'''}{147k^2}\nonumber\\
&+\frac{1096\pi\mathcal{L}'''^2}{7k}+\frac{518656\pi^2\mathcal{L}^2\mathcal{L}^{(4)}}{147k^2}+\frac{12312\pi\mathcal{L}''\mathcal{L}^{(4)}}{49k}-
\frac{4\mathcal{W}^{(4)}}{7}+\frac{6129\pi\mathcal{L}'\mathcal{L}^{(5)}}{49k}+\frac{676\pi\mathcal{L}\mathcal{L}^{(6)}}{21k}+\frac{9\mathcal{L}^{(8)}}{98}\nonumber\\
&V_0=\frac{24576000\pi^4\mathcal{L}^4\mathcal{L}'}{49k^4}-\frac{36608\pi^2\mathcal{L}\mathcal{L}'\mathcal{W}}{21k^2}+
\frac{16295680\pi^3\mathcal{L}\mathcal{L}'^3}{147k^3}-\frac{18304\pi^2\mathcal{L}^2\mathcal{W}'}{21k^2}
+\frac{29113600\pi^3\mathcal{L}^2\mathcal{L}'\mathcal{L}''}{147k^3}\nonumber\\
&-\frac{1114\pi\mathcal{W}'\mathcal{L}''}{21k}+\frac{332940\pi^2\mathcal{L}'\mathcal{L}''^2}{49k^2}-\frac{1010\pi\mathcal{L}'\mathcal{W}''}{21k}+
\frac{4272640\pi^3\mathcal{L}^3\mathcal{L}'''}{147k^3}-\frac{484\pi\mathcal{W}\mathcal{L}'''}{21k}\nonumber\\
&+\frac{737540\pi^2\mathcal{L}'^2\mathcal{L}'''}{147k^2}+\frac{41880\pi^2\mathcal{L}\mathcal{L}''\mathcal{L}'''}{7k^2}
-\frac{380\pi\mathcal{L}\mathcal{W}'''}{21k}+\frac{521440\pi^2\mathcal{L}\mathcal{L}'\mathcal{L}^{(4)}}{147k^2}
+\frac{2740\pi\mathcal{L}'''\mathcal{L}^{(4)}}{49k}\nonumber\\
&+\frac{25840\pi^2\mathcal{L}^2\mathcal{L}^{(5)}}{49k^2}+\frac{1845\pi\mathcal{L}''\mathcal{L}^{(5)}}{49k}-\frac{\mathcal{W}^{(5)}}{14}+
\frac{340\pi\mathcal{L}'\mathcal{L}^{(6)}}{21k}+\frac{180\pi\mathcal{L}\mathcal{L}^{(7)}}{49k}+\frac{5\mathcal{L}^{(9)}}{588}
\end{align}

The corresponding $G_2$ black hole with pure spin 6 hair could be
\be
a=(L_1-\mathcal{L}L_{-1}+\mathcal{W}_6 W^6_{-5})dx^++(\nu' L_{-1}+\sum_{m=-5}^{m=5} q'_{m} W^6_m)dx^-
\ee
where the independent parameters are $\mathcal{L},\mathcal{W}_6$ and $q'_5$. The other nonzero parameters are related to them by
\bea
q'_{-5}=\frac{-7 \mathcal{L}^5 q'_5 - 1032 \mathcal{L}^2 \mathcal{W}_6 q'_5}{7},\
&&\hs{3ex}q'_{-3}=5(\mathcal{L}^4 q'_5+96\mathcal{L}\mathcal{W}_6q'_5),\nonumber\\
q'_3=-5 \mathcal{L} q'_5,&&\hs{3ex}  q'_1=10 \mathcal{L}^2 q'_5,\nn\\
  q'_{-1}=-10(\mathcal{L}^3 q'_5+36\mathcal{W}_6 q'_5),\
&&\hs{3ex} \nu' = \frac{
 108000 \mathcal{W}_6 q'_5}{7}.
 \eea
 Again, we have
 \be
 q'_5=\mu_6=\frac{\a_6}{\bar{\tau}}
 \ee
 The higher spin analogue of inverse Euclidean temperature and chemical potential can be defined as
 \be
 \tau=i\frac{\gamma_1}{4\pi^2}\frac{\partial S}{\partial \mathcal{L}},\  \hs{3ex} \a_6=i\frac{\gamma_2}{4\pi^2}\frac{\partial S}{\partial\mathcal{W}_6}
 \ee
 It is easy to fix the constant
 \be
 \gamma_1=\frac{2\pi}{k},\hs{3ex} \gamma_2=\frac{7}{21600}\gamma_1.
 \ee
  With the quantity $y\equiv\frac{\mathcal{W}_6}{\mathcal{L}^3}$, the entropy should have a form $S=2\pi k \sqrt{\mathcal{L}}f(y)$. The independent holonomy equations lead to
 \bea
 1&=&f(y)^2+\frac{4(4+429y-54675y^2)f(y)'^2}{6075},\nonumber\\
 -101632 \pi^6&=&\frac{256\pi^6}{2017815046875}(2017815046875 (-397 + 4050 y)f(y)^6\nn\\&&-5380840125000 (-4 - 429 y + 54675 y^2) f(y)^5f(y)'\nn\\&&
 6643012500 (-724 - 126249 y + 4683825 y^2 + 664301250 y^3) f(y)^4f(y)'^2\nn\\&&-11809800000 ((4 + 429 y - 54675 y^2)^2)f(y)^3f(y)'^3\nn\\&&
 1458000 (-829 + 36450 y) ((4 + 429 y - 54675 y^2)^2) f(y)^2f(y)'^4\nn\\&&-2332800 (-4 - 429 y + 54675 y^2)^3 f(y)f(y)'^5+\nn\\&&
 64 (3287 + 36450 y) ((-4 - 429 y + 54675 y^2)^3) f(y)'^6).
 \eea
From the first equation, we attempt to take
\be
f(y)=\cos(\th(y))
\ee
Then
 \be
 \th(y)=\frac{1}{6}(\arcsin\frac{143}{343}-\arcsin\frac{143 - 36450 y}{343})
 \ee
 where $-\frac{4}{729}\ \leq \ y\ \leq \ \frac{1}{75}$ with two end point corresponding to the extremal black holes. It can be shown that this entropy function satisfy the second holonomy equations above.

 It is nice to see that the entropy function of the $G_2$ black hole could be written in a concise form. It is easy to see that the $G_2$ black hole share some features with the spin $\tilde 4$ black holes studied in the last section, including the form of the entropy function, the dependence on the charges, charge-conjugation violation.

\section{Black Hole with More Than One Higher Spin Hair}

In this section, we study the black hole with more than one higher spin hair in the truncated gravity with finite number of even high spins. The simplest case should be the Type I Spin $\tilde{6}$ gravity which contains spin 2, spin 4 and spin 6 hair. This could be obtained by turning off odd spin generators in $sl(6,R)$. An equally simple case is the Type II Spin $\tilde{6}$ gravity. This could be obtained by turning off odd spin generators in $sl(7,R)$. Though the spectrum of these two gravity are the same, they are different theories due to the non equivalence of $sp(6)$ and $so(7)$. Our convention for general $sl(N,R)$ algebra can be found in the Appendix C.

\subsection{Type I Spin $\tilde{6}$ Black Hole}

The type I spin $\tilde{6}$ black hole is truncated in $sl(6,R)$. The form of $a$ is
\begin{equation}
a=(L_1-\mathcal{L}L_{-1}+\sum_{s=4,6}\mathcal{W}_s W^s_{-(s-1)})dx^+ +(\nu L_{-1}+\sum_{s=4,6}\sum_{m=-(s-1)}^{m=s-1} q_{sm} W^s_m)dx^- \label{a6}
\end{equation}
and $\bar{a}$ has the similar form.
When the coefficients $\mathcal{L},\mathcal{W}_s,q_{sm}$ are constant, the equation of motion turns out to be equivalent to $[a_+,a_-]=0$.

We can work out the solution:
\begin{align}
&q_{64}=q_{62}=q_{60}=q_{6,-2}=q_{6,-4}=q_{42}=q_{40}=q_{4,-2}=0,\ \nu=\frac{144}{35} (27 \mathcal{W}_4 q_{43} + 500 \mathcal{W}_6 q_{65})\nonumber\\
&q_{63}=-5\mathcal{L}q_{65},q_{61}=10(\mathcal{L}^2q_{65}+3\mathcal{W}_4 q_{65})\nonumber\\
&q_{6,-1}=\frac{1}{3}(-30\mathcal{L}^3-230\mathcal{L}\mathcal{W}_4+120\mathcal{W}_6)q_{65}+3\mathcal{W}_4 q_{43}\nonumber\\
&q_{6,-3}=\frac{1}{3}(15\mathcal{L}^4+190\mathcal{L}^2 \mathcal{W}_4+675\mathcal{W}_4^2-160\mathcal{L}\mathcal{W}_6)q_{65}-(6\mathcal{L}\mathcal{W}_4-15\mathcal{W}_6)q_{43}\nonumber\\
&q_{6,-5}=\frac{1}{105}[(-105\mathcal{L}^5-1750\mathcal{L}^3\mathcal{W}_4-17745\mathcal{L}\mathcal{W}_4^2+1720\mathcal{L}^2\mathcal{W}_6+
49680\mathcal{W}_4\mathcal{W}_6)q_{65}\nn\\
&\hs{10ex}+(315\mathcal{L}^2\mathcal{W}_4-945\mathcal{W}_4^2-1071\mathcal{L}\mathcal{W}_6)q_{43}]\nonumber\\
&q_{41}=-3\mathcal{L}q_{43}+100\mathcal{W}_4 q_{65}\nonumber\\
&q_{4,-1}=\frac{1}{9}(-1400\mathcal{L}\mathcal{W}_4 q_{65}+3000\mathcal{W}_6)q_{65}+(3\mathcal{L}^2-17\mathcal{W}_4)q_{43}\nonumber\\
&q_{4,-3}=\frac{1}{945}[(56500\mathcal{L}^2\mathcal{W}_4-191700\mathcal{W}_4^2-175000\mathcal{L}\mathcal{W}_6)q_{65}+
(-945\mathcal{L}^3+11781\mathcal{L}\mathcal{W}_4+31500\mathcal{W}_6)q_{43}].\nn
\end{align}
We can write $q_{43}=\frac{\a_4}{\bar{\tau}},q_{65}=\frac{\a_6}{\bar{\tau}}$, where $\a_4$ and $\a_6$ are spin-4 and spin-6 chemical potentials,  defined as:
\begin{equation}
\a_4=\frac{ic_2}{4\pi^2}\frac{\partial S}{\partial \mathcal{W}_4},\hs{3ex}\a_6=\frac{ic_3}{4\pi^2}\frac{\partial S}{\partial\mathcal{W}_6}. \label{mu46}
\end{equation}
And $\tau=\frac{ic_1}{4\pi^2}\frac{\partial S}{\partial\mathcal{L}}$ is analogue of the inverse Euclidean temperature of BTZ black hole. The coefficients can be fixed by comparing with Ward identity:
\be
c_1=\frac{2\pi}{k},\hs{3ex}c_2=\frac{35}{1296}c_1,\hs{3ex}c_3=\frac{7}{2880}c_1.
 \ee

 We assume the entropy $S$ has the form $2\pi k\sqrt{\mathcal{L}}f(y,z)$, where $y=\mathcal{W}_4/\mathcal{L}^2,z=\mathcal{W}_6/\mathcal{L}^3$ are dimensionless parameters. Then we have
\begin{align}
&\tau=\frac{i}{2\sqrt{\mathcal{L}}}(f-4y\partial_y f-6z\partial_z f)\nonumber\\
&\a_4=\frac{35i}{1296\mathcal{L}^{3/2}}\partial_y f\nonumber\\
&\a_6=\frac{7i}{2880\mathcal{L}^{5/2}}\partial_z f.
\end{align}

 The holonomy is just $\omega=2\pi(\tau a_+ -\bar{\tau}a_-)$. In spin-$N$ gravity the holonomy $\omega$ has eigenvalues $(N-1)i\pi,(N-3)i\pi,\dots,-(N-1)i\pi$. In our $N=6$ case, the holonomy condition gives the following constraints
\begin{eqnarray}
tr(\omega)=tr(\omega^3)=tr(\omega^5)=0,\hs{3ex}tr(\omega^2)=-70\pi^2,\nn\\
tr(\omega^4)=1414\pi^4,\hs{3ex}tr(\omega^6)=-32710\pi^6.
\end{eqnarray}
The first three equations are automatically satisfied since $tr(V^{s_1}_{m_1}\dots V^{s_n}_{m_n})$ can be non-zero only when $m_1+\dots+m_n=0$($V^s_m$ can be $L_m$ or $W^4_m$ or $W^6_m$), and in our solution we have only $V^s_m$ with odd $m$. So all the odd equations are trivially satisfied. In principle we can solve the later three equations to get the expression of $f(y,z)$. But they're non-linear partial differential equations and are hard to solve. And the full representation of their explicit form is not quite illustrative. What we can do is to expand $f(y,z)=1+C_{1,0}y+C_{0,1}z+\dots$ and obtain a few coefficients perturbatively.

\subsection{Type II Spin $\tilde{6}$ Black Hole}

The type II spin $\tilde{6}$ black hole is truncated in $sl(7,R)$. The form of $a$ is the same as (\ref{a6}):
\begin{equation}
a=(L_1-\mathcal{L}L_{-1}+\sum_{s=4,6}\mathcal{W}_s W^s_{-(s-1)})dx^+ +(\nu L_{-1}+\sum_{s=4,6}\sum_{m=-(s-1)}^{m=s-1} q_{sm} W^s_m)dx^-.
\end{equation}
 Now the subalgebra that consists of $L_1,L_0,L_{-1},W^4_3,\dots,W^4_{-3},W^6_5,\dots,W^6_{-5}$ is $so(7,R)$, so is different from $sp(6,R)$, which is the algebra of type I spin $\tilde{6}$ gravity. The solution of $[a_+,a_-]=0$ is:
\begin{align}
&q_{64}=q_{62}=q_{60}=q_{6,-2}=q_{6,-4}=q_{42}=q_{40}=q_{4,-2}=0,\hs{2ex}\nu = \frac{540}{7} (3 q_{43} W_4 + 200 q_{65} W_6), \nonumber\\
&q_{63}=-5\mathcal{L}q_{65},q_{61}=10\mathcal{L}^2q_{65},q_{6,-1}=(-10\mathcal{L}^3-360q_{65}\mathcal{W}_6)q_{65}+3\mathcal{W}_4q_{43}\nonumber\\
&q_{6,-3}=(5\mathcal{L}^4+540\mathcal{W}_4^2+480\mathcal{L}\mathcal{W}_6)q_{65}-6\mathcal{L}\mathcal{W}_4q_{43}\nonumber\\
&q_{6,-5}=\frac{1}{7}(-7\mathcal{L}^5-3108\mathcal{L}\mathcal{W}_4^2-1032\mathcal{L}^2\mathcal{W}_6)q_{65}+(3\mathcal{L}^2\mathcal{W}_4-
18\mathcal{W}_4^2)q_{43}\nonumber\\
&q_{41}=360\mathcal{W}_4 q_{65}-3\mathcal{L}q_{43},q_{4,-1}=(-3\mathcal{L}^2-30\mathcal{W}_4)q_{43}-560\mathcal{L}\mathcal{W}_4 q_{65}\nonumber\\
&q_{4,-3}=(-\mathcal{L}^3+22\mathcal{L}\mathcal{W}_4+120\mathcal{W}_6)q_{43}+\frac{1}{21}(4520\mathcal{L}^2\mathcal{W}_4-30240\mathcal{W}_4^2) q_{65}.
\end{align}

Similarly, we can write $q_{43}=\frac{\a_4}{\bar{\tau}},q_{65}=\frac{\a_6}{\bar{\tau}}$, with $\a_4,\a_6$ being spin 4 and spin 6 chemical potentials defined by (\ref{mu46}). But now
\be
c_1=\frac{2\pi}{k},\hs{3ex}c_2=\frac{7}{540}c_1,\hs{3ex}c_3=\frac{7}{21600}c_1.
\ee
 The holonomy conditions in this case are:
\begin{align}
&tr(\omega)=tr(\omega^3)=tr(\omega^5)=tr(\omega^7)=0,\nonumber\\
&tr(\omega^2)=-112\pi^2,tr(\omega^4)=3136\pi^4,tr(\omega^6)=-101632\pi^6.
\end{align}
Similar to the Type I spin ${\tilde 6}$ black hole, the entropy function cannot be worked out in an exact form. We are not going to the details anymore.

\section{Conclusion and Discussions}

In this paper we investigated the black holes in truncated higher spin AdS$_3$ gravity theories. As we are  interested in the black hole solutions with one single higher spin hair, we mainly focused on the higher spin gravity theory with gauge group of rank two. We found that besides the known spin-3 gravity on $sl(3,R)$ algebra, there are only two other theories: the spin $\tilde 4$ gravity and $G_2$ gravity. In the former case, the theory could be related to the Vasiliev's higher spin gravity by truncations in two different ways, with its asymptotical symmetry algebra being $W(2,4)$. In the latter case, the theory is defined with the gauge group $G_2$, and its asymptotical symmetry algebra is $W(2,6)$. It is remarkable that the non-existence of other higher spin gravity with single higher spin hair is in accordance with the fact that there is no generic $W(2,s)$-algebra with $s>6$. In many aspects, the spin $\tilde 4$ gravity and $G_2$ gravity are as simple as the spin 3 gravity.

In both spin $\tilde 4$ gravity and $G_2$ gravity, we constructed the exact black hole solutions with higher spin hair successfully. It is nice to see that in both cases the entropy functions could be written in an exactly closed form. This nice feature allows us to take these two black holes along with the spin 3 black hole as the prototypes to test various ideas on higher spin gravity. One remarkable feature of our black holes is that they are sensitive to the sign of higher spin charges. This is very different from the spin 3 black hole or usual $U(1)$ charged  black holes. It would be interesting to understand this fact from other points of view.

 It is a distinguishable feature of AdS$_3$ higher spin gravity to allow truncation to a theory of finite number of higher spin fields, since there is no finite truncation in $d>3$ AdS$_d$ spacetime. As shown in Section 2, besides the original $sl(N,R)$ higher spin gravity, there are two other kinds of finite truncations, Type I Spin $\widetilde{2N}$ and Type II Spin $\widetilde{2N}$ gravity theories, based on the $sp(2N,R)$ and $so(2N+1,R)$ Lie algebra respectively. However, these are not the only possibilities. As shown in Section 4, there exists a higher spin gravity based on the exceptional algebra $G_2$. The $G_2$ gravity cannot be obtained directly from Vasiliev's theory by simply doing even spin truncation. Nevertheless, it is closely related to  Type II Spin $\tilde{6}$ gravity. Actually, by turning off all the spin 4 generators in  Type II Spin $\tilde{6}$ gravity, we obtain a well-defined $G_2$ gravity. This raises an interesting question if the Chern-Simons gravity based on other Lie algebras can give us consistent higher spin gravity.


Another interesting application of our construction is to embed the spin 4 black hole into Vasiliev's theory on $\widetilde{hs}(\l)$ algebra. In this case, as the spin 4 field is the lowest high spin, it will generate other higher even spins in the theory. In principle, the partition function could be computed perturbatively  and compared with the CFT study\cite{Ah,Gaberdial2:2011}. This will provide another tractable example to test HS/CFT correspondence.


The higher spin topologically massive gravity(HSTMG) was constructed in \cite{Spin-3 TMG:2011,Sahoo 1:2011,Sahoo 2:2011,Hstmg}. The truncated version of HSTMG is obvious.  Though the physical meaning of the general solution of HSTMG is not clear so far\footnote{For the first step to this problem, see\cite{Solution}},  the black holes in higher spin gravity is naturally identified with the black holes in HSTMG  since every solution of higher spin gravity is automatically the solution of HSTMG. This do not mean that the physical quantity is the same as before. The Chern-Simons level do have effect in the quantity such as entropy. A naive guess is that we should replace the level $k\to(1-\frac{1}{\mu})k$ (or $(1+\frac{1}{\mu})k$) everywhere to get the correct answer. We leave this problem to further study.

\vspace*{10mm}
\noindent {\large{\bf Acknowledgments}}

The work was in part supported by NSFC Grant No.~10975005, ~11275010.

\vspace*{5mm}

\section*{Appendix A: Convention on $sp(4,R)$}
The explicit forms of the generators $L_1,L_0,L_{-1},W^4_3,\dots,W^4_{-3}$ are:
\\
\ \ \ $
L_1=\left(
\begin{array}{cccc}
0&0&0&0\\-\sqrt{3}&0&0&0\\0&-2&0&0\\0&0&-\sqrt{3}&0
\end{array}
\right)
$,$L_{-1}=\left(\begin{array}{cccc}
0&\sqrt{3}&0&0\\0&0&2&0\\0&0&0&\sqrt{3}\\0&0&0&0
\end{array}\right)$,$
L_0=\frac{1}{2}\left(\begin{array}{cccc}
3&0&0&0\\0&1&0&0\\0&0&-1&0\\0&0&0&-3
\end{array}\right)$ \\
$W^{4}_3=\left(
\begin{array}{cccc}
0&0&0&0\\0&0&0&0\\0&0&0&0\\-6&0&0&0
\end{array}
\right)$,$W^4_2=\left(
\begin{array}{cccc}
0&0&0&0\\0&0&0&0\\ \sqrt{3}&0&0&0\\0&-\sqrt{3}&0&0
\end{array}
\right)$,$W^4_1=\frac{1}{5}\left(
\begin{array}{cccc}
0&0&0&0\\-2\sqrt{3}&0&0&0\\0&6&0&0\\0&0&-2\sqrt{3}&0
\end{array}
\right)$\\
$W^4_0=\frac{3}{10}\left(
\begin{array}{cccc}
1&0&0&0\\0&-3&0&0\\0&0&3&0\\0&0&0&-1
\end{array}
\right)$,$W^4_{-1}=\frac{1}{5}\left(
\begin{array}{cccc}
0&2\sqrt{3}&0&0\\0&0&-6&0\\0&0&0&2\sqrt{3}\\0&0&0&0
\end{array}
\right)$,\\$W^4_{-2}=\left(
\begin{array}{cccc}
0&0&\sqrt{3}&0\\0&0&0&-\sqrt{3}\\0&0&0&0\\0&0&0&0
\end{array}
\right)$,$W^4_{-3}=\left(
\begin{array}{cccc}
0&0&0&6\\0&0&0&0\\0&0&0&0\\0&0&0&0
\end{array}
\right) $

The commutation relations are
\begin{align}
&[L_i,L_j]=(i-j)L_{i+j},\hs{2ex}[L_i,W^4_m]=(3i-m)W_{i+m}\nonumber\\
&[W^4_3,W^4_0]=\frac{3}{5}W^4_3,\hs{2ex}[W^4_3,W^4_{-1}]=\frac{12}{5}W^4_2,\hs{2ex}[W^4_3,W^4_{-2}]=6W^4_1+\frac{18}{5}L_1\nonumber\\
&[W^4_3,W^4_{-3}]=12W^4_0+\frac{108}{5}L_{0},\hs{2ex}[W^4_2,W^4_1]=-\frac{2}{5}W^4_3,\hs{2ex}[W^4_2,W^4_0]=-\frac{3}{5}W^4_2\nonumber\\
&[W^4_2,W^4_{-1}]=-\frac{6}{5}L_1,\hs{2ex}[W^4_2,W^4_{-2}]=2W^4_0-\frac{12}{5}L_0,\hs{2ex}[W^4_2,W^4_{-3}]=6W^4_{-1}+\frac{18}{5}L_{-1}\nonumber\\
&[W^4_1,W^4_0]=-\frac{3}{5}W^4_1+\frac{18}{25}L_1,\hs{2ex}[W^4_1,W^4_{-1}]=-\frac{4}{5}W^4_0+\frac{12}{25}L_0,\hs{2ex}[W^4_1,W^4_{-2}]=-\frac{6}{5}L_{-1}\nonumber\\
&[W^4_1,W^4_{-3}]=\frac{12}{5}W^4_{-2},\hs{2ex}[W^4_0,W^4_{-1}]=-\frac{3}{5}W^4_{-1}+\frac{18}{25}L_{-1},\hs{2ex}[W^4_0,W^4_{-2}]=-\frac{3}{5}W^4_{-2}\nonumber\\
&[W^4_0,W^4_{-3}]=\frac{3}{5}W^4_{-3},\hs{2ex}[W^4_{-1},W^4_{-2}]=-\frac{2}{5}W^4_{-3}.
\end{align}
and the others are zero. The first line of these commutation relations simply means that the spectrum of this theory is spin 2 and spin 4.

\section*{Appendix B: Convention to $G_2$}
The explicit forms of generators $L_1,L_0,L_{-1},W^6_{-5},\cdots,W^6_5$ are
\\
 $
L_1=\left(
\begin{array}{ccccccc}
0&0&0&0&0&0&0\\-\sqrt{6}&0&0&0&0&0&0\\0&-\sqrt{10}&0&0&0&0&0\\0&0&-2\sqrt{3}&0&0&0&0\\0&0&0&-2\sqrt{3}&0&0&0\\0&0&0&0&-\sqrt{10}&0&0\\0&0&0&0&0&-\sqrt{6}&0
\end{array}
\right)$\\
$L_0=\left(\begin{array}{ccccccc}
3&0&0&0&0&0&0\\0&2&0&0&0&0&0\\0&0&1&0&0&0&0\\0&0&0&0&0&0&0\\0&0&0&0&-1&0&0\\0&0&0&0&0&-2&0\\0&0&0&0&0&0&-3\end{array}\right)
$,
$
L_{-1}=\left(
\begin{array}{ccccccc}
0&\sqrt{6}&0&0&0&0&0\\0&0&\sqrt{10}&0&0&0&0\\0&0&0&2\sqrt{3}&0&0&0\\0&0&0&0&2\sqrt{3}&0&0\\0&0&0&0&0&\sqrt{10}&0\\0&0&0&0&0&0&\sqrt{6}\\0&0&0&0&0&0&0\end{array}\right)
$\\
$W^6_5=\left(\begin{array}{ccccccc}
0&0&0&0&0&0&0\\0&0&0&0&0&0&0\\0&0&0&0&0&0&0\\0&0&0&0&0&0&0\\0&0&0&0&0&0&0\\-120\sqrt{6}&0&0&0&0&0&0\\0&-120\sqrt{6}&0&0&0&0&0\end{array}\right)
$,
$W^6_4=\left(\begin{array}{ccccccc}
0&0&0&0&0&0&0\\0&0&0&0&0&0&0\\0&0&0&0&0&0&0\\0&0&0&0&0&0&0\\24\sqrt{15}&0&0&0&0&0&0\\0&0&0&0&0&0&0\\0&0&-24\sqrt{15}&0&0&0&0\end{array}\right)$
\\
$W^6_3=\left(\begin{array}{ccccccc}
0&0&0&0&0&0&0\\0&0&0&0&0&0&0\\0&0&0&0&0&0&0\\-16\sqrt{5}&0&0&0&0&0&0\\0&8\sqrt{10}&0&0&0&0&0\\0&0&8\sqrt{10}&0&0&0&0\\0&0&0&-16\sqrt{5}&0&0&0\end{array}\right)$,
\\$W^6_2=\left(\begin{array}{ccccccc}
0&0&0&0&0&0&0\\0&0&0&0&0&0&0\\4\sqrt{
  15}&0&0&0&0&0&0\\0&-4\sqrt{
  30}&0&0&0&0&0\\0&0&0&0&0&0&0\\0&0&0&4\sqrt{30}&0&0&0
\\0&0&0&0&-4\sqrt{15}&0&0\end{array}\right)$\\
$W^6_1=\left(\begin{array}{ccccccc}
0&0&0&0&0&0&0\\-\frac{20\sqrt{6}}{7}&0&0&0&0&0&0\\0&\frac{36\sqrt{10}}{7}&0&0&0&0&0\\0&0&-\frac{40\sqrt{3}}{7}&0&0&0&0\\0&0&0&-\frac{40\sqrt{3}}{7}&0&0&0\\0&0&0&0&\frac{36\sqrt{10}}{7}&0&0\\0&0&0&0&0&-\frac{20\sqrt{6}}{7}&0\end{array}\right)$
\\$W^6_0=\left(\begin{array}{ccccccc}
\frac{20}{7}&0&0&0&0&0&0\\0&-\frac{80}{
  7}&0&0&0&0&0\\0&0&\frac{100}{7}&0&0&0&0\\0&0&0&0&0&0&
0\\0&0&0&0&-\frac{100}{
  7}&0&0\\0&0&0&0&0&\frac{80}{7}&0\\0&0&0&0&0&0&-\frac{20}{7}\end{array}\right)$
  \\and
  \be
  W^6_{-m}=(-1)^m (W^6_m)^{\dagger}\label{al}
  \ee
  for $m=1,2,\cdots,5$.

The commutation relations are
\begin{align}
&[L_i,L_j]=(i-j)L_{i+j},\ [L_i,W^6_m]=(5i-m)W^6_{m+i}\nonumber\\
&[W^6_5,W^6_{-5}]= -2160 W^6_0+\frac{216000}{7}L_0,\  [W^6_5,W^6_{-4}]=-1080 W^6_1+\frac{21600}{7}L_1\nonumber\\
&[W^6_5,W^6_{-3}]=-480W^6_2,\ [W^6_5, W^6_{-2}]=-180 W^6_3\nonumber\\
&[W^6_5,W^6_{-1}]=-\frac{360}{7}W^6_4,\ [W^6_5,W^6_0]=-\frac{60}{7}W^6_5\nonumber\\
&[W^6_4,W^6_{-4}]=-432W^6_0-\frac{17280}{7} L_0,\ [W^6_4,W^6_{-3}]=-120W^6_1 -\frac{4320}{7} L_1\nonumber\\
&[W^6_4,W^6_{-2}]=0,\ [W^6_4,W^6_{-1}]=\frac{180}{7}W^6_3\nonumber\\
&[W^6_4,W^6_0]=\frac{120}{7}W^6_4,\ [W^6_4,W^6_1]=\frac{36}{7}W^6_5\nonumber\\
&[W^6_3,W^6_{-3}=16 W^6_0 + \frac{2880}{7} L_0,\ [W^6_3,W^6_{-2}]=40 W^6_1 + \frac{1440}{7} L_1\nonumber\\
&[W^6_3,W^6_{-1}]=\frac{160}{7}W^6_2,\ [W^6_3,W^6_0]=\frac{20}{7}W^6_3\nonumber\\
&[W^6_3,W^6_1]=-\frac{40}{7}W^6_4,\ [W^6_3,W^6_2]=-4W^6_5\nonumber\\
&[W^6_2,W^6_{-2}]=24 W^6_0-\frac{720}{7} L_0,\ [W^6_2,W^6_{-1}]=-\frac{720}{7}L_1\nonumber\\
&[W^6_2,W^6_0]=-\frac{80}{7}W^6_2,\ [W^6_2,W^6_1]=-\frac{60}{7}W^6_3\nonumber\\
&[W^6_1,W^6_{-1}]=-\frac{96}{7} W^6_0 + \frac{1440}{49} L_0,\ [W^6_1,W^6_0]=-\frac{80}{7} W^6_1 + \frac{3600}{49} L_1
\end{align}
and the other commutations can be derived using (\ref{al}) or simplify to be 0.
Note that the first line means that the spectrum contains spin 2 and spin 6.
\section*{Appendix C: General Lie algebra generators and commutation rules in $sl(N,R)$}

Here we use this set of $L_0,L_1,L_{-1}$ to generate the whole set of generators:

\begin{align}
&L_0=\frac{1}{2}\left(\begin{array}{ccccc}
N-1&0&\dots&0&0\\0&N-3&\dots&0&0\\ \vdots&\vdots&\ddots&\vdots&\vdots\\0&0&\dots&-(N-3)&0\\0&0&\dots&0&-(N-1)
\end{array}\right)\nonumber\\
&L_1=\left(
\begin{array}{ccccc}
0&0&\dots&0&0\\-\sqrt{N-1}&0&\dots&0&0\\0&-\sqrt{2N-4}&\dots&0&0\\ \vdots&\vdots&\ddots&\vdots&\vdots\\0&0&\dots&-\sqrt{N-1}&0
\end{array}
\right)\nonumber\\
&L_{-1}=\left(\begin{array}{ccccc}
0&\sqrt{N-1}&0&\dots&0\\0&0&\sqrt{2N-4}&\dots&0\\ \vdots&\vdots&\vdots&\ddots&\vdots\\0&0&0&\dots&\sqrt{N-1}\\0&0&0&\dots&0
\end{array}\right)
\end{align}
Here $L_{1(i+1,i)}=-\sqrt{Ni-i^2}$, $L_{-1(i,i+1)}=\sqrt{Ni-i^2}$, and they satisfy commutation relation $[L_i,L_j]=(i-j)L_{i+j}$.

The other generators are denoted by $W^s_m$($3\leq s\leq N, -(s-1)\leq m\leq(s-1)$), and are derived out by
\begin{equation}
W^s_{s-1}=L_1^{s-1},\hs{2ex}W^s_{m-1}=-\frac{1}{s+m-1}[L_{-1},W^s_m].
\end{equation}

 \end{document}